# Optimization and Performance Evaluation of $Cs_2CuBiCl_6$ Double Perovskite Solar Cell for Lead-Free Photovoltaic Applications

[1]Syeda Anber Urooj Wasti[1], Sundus Naz[1], Ammara Sattar[2], Asad Yaqoob[3], Ateeq ul Rehman[4], Fawad Ali[5*], Shahbaz Afzal[6*]

[1]Department of Physics, University of Agriculture Faisalabad.
[2]Department of Physics, Faculty of Engineering and Applied Sciences (FEAS), Riphah International University, Islamabad, 44000, Pakistan.
[3]Department of Physics, Abbottabad University, Pakistan.
[4]Institute of Physics, Baghdad ul Jadeed Campus, The Islamia University of Bahawalpur, Bahawalpur, 63100, Pakistan.
[5]Department of Physics, University of Education Lahore, DG Khan Campus, Punjab, Pakistan.
[6]Nanophotonics Research Centre, Shenzhen University and Key Laboratory of Optoelectronic Devices and Systems of Ministry of Education and Guangdong Province, College of Optoelectronics Engineering, Shenzhen University, Shenzhen 518060, China.

*Corresponding author(s): Fawad Ali, Email: fawadali2022@163.com; Shahbaz Afzal, Email: shahbazafzal2216@gmail.com.*

**Abstract**

In the previous decade, there has been a significant advancement in the performance of perovskite solar cells (PSCs), characterized by a notable increase in efficiency from 3.8% to 25%. Nonetheless, PSCs face many problems when we commercialize them because of their toxicity and stability. Consequently, lead-PSCs need an alternative solar cell with high performance and low processing cost; lead-free inorganic perovskites have been explored. Recent research showcased Cs2CuBiCl6, a lead-free inorganic double perovskite material with remarkable photoelectric characteristics and exceptional environmental robustness. To investigate the potential of $Cs_2CuBiCl_6$ material, the solar cell structure FTO/ETL/$Cs_2CuBiCl_6$/HTL/Au was used and analyzed through a solar cell capacitance simulator (SCAPS-1D). $CeO_2$ is used as the Electron transport layer (ETL), and CuI is the Hole transport layer (HTL). Furthermore, the research examined the optimization of different parameters of the absorber layer (AL), such as thickness, defect density, electron affinity, band gap, and operational temperature. In the end, it has been noticed that by setting the temperature at 300 K and an electron affinity of 4.3 eV of the absorber layer, the PSCs achieve the highest efficiency of 24.51 %, FF of 43.01 %, $V_{oc}$ of 1.73V, and $J_{sc}$ of 32.82mA/cm$^2$. This is the highest $Cs_2CuBiCl_6$ double PSCs efficiency we've reached yet. In theoretical studies, 17.03% of PCE was achieved using $Cs_2CuBiCl_6$ as an active layer. The analysis underscores the significant potential of $Cs_2CuBiCl_6$ as an absorbing layer in developing highly efficient lead-free all-inorganic PSCs.

1. **Introduction**

Energy consumption is currently at its highest levels, with demand projected to rise further due to the ever-increasing global population. This situation underscores the urgent need to identify alternative energy sources beyond fossil fuels. Such efforts are essential for sustaining a healthy society and environment and meeting global energy requirements. Among the various solutions, renewable energy-based solar cell technologies have emerged as one of the most promising approaches to address the ongoing energy crisis [1, 2]. Since the 1950s, silicon- as the primary material in first-generation solar cells- has been widely used in photovoltaic applications due to its high efficiency. However, its extraction and production costs remain prohibitively expensive [1]. A first-generation solar cell offers superior efficiency but comes with higher production costs. Second-generation solar cells, also known as thin-film solar cells, were introduced in response to these challenges. These are fabricated using amorphous silicon, cadmium telluride (CdTe), and copper indium gallium Selenide (CIGS). While second-generation solar cells are lighter, more flexible, and cheaper to produce than first-generation silicon-based solar cells, they generally exhibit lower efficiencies, shorter lifespans, and pose potential environmental and health risks due to using toxic materials during production and disposal [2]. The third generation of perovskite solar cells (PSCs) was developed to overcome these limitations. PSCs have garnered significant attention over the past two decades due to their rapid efficiency improvements and potential for low-cost production [3]. Several unique physic-chemical properties contribute to the popularity of PSCs: i) an adjustable energy band gap, ii) efficient utilization of incident light utilization, iii) cost-effective and versatile preparation techniques, and iv) rapid progress in research and development. Remarkably, PSCs achieved a dramatic enhancement in power conversion efficiency (PCE), increasing from 3.8% to 25.7% within a short time period [4, 5]. Structurally, PSCs typically comprise a transparent conducting oxide TCO (e.g., FTO and ITO), an electron transport layer ETL, a hole transport layer HTL, and a perovskite layer (i.e., perovskite layer positioned between the ETL and HTL layer. Perovskites are materials characterized by the general formula $ABX_3$, where 'A' represents a large cation, 'B' denotes a smaller Cation, and 'X' is a halide anion. By introducing new elements through doping, novel perovskite structures with enhanced optoelectronic properties can be engineered. Bismuth-based double perovskite is a particularly intriguing class of perovskite with researchers' attention. These materials offer significant potential for solar energy applications due to their high stability, efficient generation of electron-hole pairs, and tunable optoelectronic properties [6, 7]. Double perovskite (DP), also known as an elpasolite, is derived by replacing one monovalent $B^+$ and one trivalent $B^{3+}$ ion with a $Pb^{2+}$ ion in the traditional perovskite structure, resulting in the formula $A_2B^+B^{3+}X_6(Cs_2CuBiCl_6)$ [8]. In Double perovskite solar cells, the absorber layer- a thin film of double perovskite material- is crucial for absorbing light and generating electron-hole pairs. These absorber layers are central to distinct characteristics and enhanced performance of double perovskite solar cells. Compared to conventional perovskites, double perovskites offer notable advantages, including higher stability, superior electron-hole pair production, and customizable optoelectronic properties, making them a promising alternative for next-generation photovoltaic technologies.

$Cs_2CuBiCl_6$ is a promising lead-free double perovskite due to its optimal 1.1 eV band gap, facilitating efficient light absorption and conversion into electricity while exhibiting impressive stability for sustained performance. This exceptional stability enhances the material's suitability for solar cell applications [9]. Moreover, its remarkable optical properties further augment its ability to effectively capture and convert solar energy [10]. Double perovskites offer a sustainable alternative by replacing the lead with a combination of other metals. For instance, one of the most researched double perovskites is $Cs_2AgBiBr_6$, where silver (Ag) and bismuth (Bi) substitute lead, resulting in a lead-free material with promising photovoltaic properties [11]. Over the past few years, extensive research has focused on developing double perovskite to improve overall stability and efficiency while maintaining a low-cost and lead-free solar cell system. Recent studies have demonstrated the potential of double perovskite in photovoltaic applications. For example, Kale et al. reported a power conversion efficiency (PCE) of 17.03% using $TiO_2$ as ETL $Cu_2O$ as HTL and $Cs_2CuBiCl_6$ as absorber layer [12]. Mohandas et al. utilized $Cs_2AgBiBr_6$, as the active layer and reported a PCE efficiency of 1.44% by using SCAPS-1D. Furthermore, Ferdous et al. simulated $Cs_2TiI_2Br_4$-based double PSCs and gained an impressive efficiency of 23.41% [13]. In this research, $Cs_2CuBiCl_6$ was selected as the absorber layer due to its superior material capabilities and the scarcity of comparative studies on this compound. According to the literature, only one study has been conducted using $Cs_2CuBiCl_6$ as a double perovskite layer [12].

This work aims to address this gap by systematically exploring the photovoltaic performance of lead-free double perovskite solar cells employing $FTO/CeO_2/Cs_2CuBiCl_6/CuI/Au$ configuration using SCAPS-1D simulation software. In the proposed structure, $Cs_2CuBiCl_6$ serves as double perovskite layer (DPL), $CeO_2$ as the electron transport layer (ETL), CuI as the hole transport layer (HTL), FTO as the transparent conductive substrate, and Au as the back-contact material. The significance of this work lies in the detailed investigation of factors influencing the performance of $Cs_2CuBiCl_6$-based PSCs. Key parameters of the absorber layer, including thickness, defect density; electron affinity and operating temperature were systematically analyzed to optimized device performance. The impacts of these parameters on power conversion efficiency (PCE), fill factor (FF), open-circuit voltage (Voc), and short-circuit current density (Jsc) were comprehensively evaluated [14]. In the presented $FTO/CeO_2/Cs_2CuBiCl_6/CuI/Au$ solar cell structure, $CeO_2$ and CuI are chosen as the electron transport layer (ETL) and hole transport layer (HTL), respectively, due to their unique properties that enhance device performance. $CeO_2$, an n-type semiconductor with a wide band gap of 3.2 eV, ensures minimal parasitic absorption and efficient light transmission to the absorber layer. Its high electron mobility facilitates rapid extraction and transport of photogenerated electrons, reducing recombination losses and improving efficiency. Additionally, $CeO_2$ is thermally and chemically stable, making it suitable for long-term operational stability in solar cells. Similarly, CuI, as a p-type semiconductor, provides excellent hole mobility (up to 44 cm²/V·s), ensuring efficient hole extraction and transport. Its wide band gap (3.1 eV) minimizes optical losses, while its low cost and easy fabrication makes it suitable for scalable applications. The energy level alignment of CuI with $Cs_2CuBiCl_6$ enables efficient hole transfer and suppresses electron backflow, enhancing power conversion efficiency. Together, $CeO_2$ and CuI create a synergistic charge transport framework that reduces recombination, optimizes carrier collection, and ensures high stability, making them ideal for sustainable, high-performance solar cells. However, through the optimization of critical parameters such as absorber layer thickness, electron affinity, band gap, defect density, electron affinity, this research demonstrates and able to obtained the PCE of 20.43% which is greater than theoretical work (17.03%) performed by A.J. Kale et al with same double perovskite layer [12]. Table 3 presents previously reported work on different double perovskite layers and present work. By focusing on modeling and optimizing these factors, this study contributes to developing more efficient, stable, and sustainable photovoltaic technologies. The findings provide valuable insights into the design and optimization of lead-free double perovskite solar cells, addressing the global demand for renewable energy solutions while advancing the frontiers of perovskite solar cell technology.

2. **Device Simulation**

Simulation provides an effective and reliable approach to analyzing the intrinsic physical properties of a solar cell, eliminating the need for complex and resource-intensive laboratory fabrication. Through simulation, the physical characteristics of solar cells can be accurately assessed while significantly reducing material while significantly reducing materials consumption, financial cost, and time [15]. To evaluate the performance of PCS, various modeling software tools, including GPVDM, SILVACO, AMPS, ATLAS, COMSOL, and SCAPS, are widely used [16]. SCAPS-1D was employed to simulate and analyze the performance of the FTO/CeO$_2$/Cs$_2$CuBiCl$_6$/CuI/Au solar cell configuration. SCAPS-1D is particularly advantageous due to its user-friendly interface and capability to perform simulations under light and dark conditions. It enables the evaluation of key solar cell properties, including current-voltage (I-V) characteristics, capacitance-voltage (C-V) profile, capacitance-frequency (C-f) behavior, and quantum efficiency (QE). Simulations were conducted under standard AM 1.5G illumination and at a temperature of 300 Kelvin. The software allows for comprehensive AC and DC electrical calculations in varying temperature and illumination conditions, providing flexibility and precision in solar cell modeling [21]. Voltage readings were calibrated to a zero-volt reference, and the frequency for AC calculations was fixed at 1 MHz [21]. Figure 1 (a) illustrates the presented solar cell configuration, which includes a layered configuration of FTO/CeO$_2$/Cs$_2$CuBiCl$_6$/CuI/Au, where Cs$_2$CuBiCl$_6$ serves as the absorber layer, CeO$_2$ acts as the electron transport layer (ETL), and CuI is employed as the hole transport layer (HTL). The absorber layer is positioned at the top of the cell, with the p-type CuI and n-type CeO$_2$ layers placed beneath it [17]. Tables 1 and 2 detail the simulation parameters, carefully selected based on experimental data and previous simulation studies [12, 18, 19]. The simulation incorporated physical properties such as carrier transport processes, electromagnetic field distribution, recombination characteristics, and specific current densities. SCAPS-1D operates by solving Poisson's equation alongside the coupled continuity equations for electrons and holes, providing a robust framework for modeling the physical and electrical behavior of the device as shown in equations 1, 2, and 3 [20, 21].

$$\frac{d}{dx}\left(\varepsilon(\chi)\frac{d\psi}{dx}\right) = q[p(x) - n(x) + N_D^+(x) - N_A^-(x) + p_t(x) - n_t(x)] \quad (1)$$

$$\frac{\frac{1}{J}\partial J_p}{\partial x} + R_p(x) - G(x) = 0 \quad (2)$$

$$-\frac{1}{J}\partial J_n/\partial x + R_n(x) - G(x) = 0 \quad (3)$$

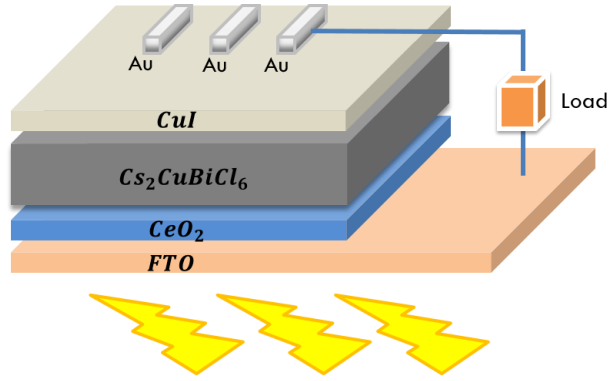

**Figure 1. (a)** Schematic Diagram of the Presented Solar Cell.

**Table 1.** Numerical Parameters for the device designed in this study.

| Thickness (μm) | 0.1 | 0.45 | 0.45 | 0.4 |
|---|---|---|---|---|
| Electron affinity (eV) | 2.1 | 4 | 4.6 | 4.3 |
| Band gap (eV) | 3.1 | 1.21 | 3.5 | 3.5 |
| CB effective density of states (cm$^{-3}$) | $2.80\times10^{19}$ | $1.00\times10^{18}$ | $1.00\times10^{20}$ | $2.20E\times10^{18}$ |
| VB effective density of states (cm$^{-3}$) | $1.00\times10^{19}$ | $1.00\times10^{18}$ | $2.00\times10^{21}$ | $1.80\times10^{19}$ |
| Dielectric permittivity (eV) | 6.5 | 3.72 | 9 | 9 |
| Electron mobility, μn (cm$^2$/Vs) | $1.00\times10^{2}$ | $2.00\times10^{0}$ | $1.00\times10^{2}$ | $2.00\times10^{1}$ |
| Hole mobility, μh (cm$^2$/Vs) | $4.39\times10^{1}$ | $2.00\times10^{0}$ | $2.50\times10^{1}$ | $1.00\times10^{1}$ |
| Uniform donor density $N_D$ (cm$^{-3}$) | 0 | $1.00\times10^{13}$ | $1.00\times10^{21}$ | $1.00\times10^{18}$ |
| Uniform acceptor density $N_A$ (cm$^{-3}$) | $1.00\times10^{18}$ | $1.00\times10^{17}$ | 0 | 0 |
| Electron thermal velocity (cm/s) | $1.00\times10^{7}$ | $2.00\times10^{0}$ | $1.00\times10^{7}$ | $1.00\times10^{7}$ |
| Hole thermal velocity (cm/s) | $1.00\times10^{7}$ | $2.00\times10^{0}$ | $1.00\times10^{7}$ | $1.00\times10^{7}$ |
| Defect density $N_t$ (cm$^{-3}$) | $1.00\times10^{15}$ | $1.00\times10^{15}$ | $1.00\times10^{15}$ | $1.00\times10^{15}$ |
| References | [18] | [12] | [18] | [22] |

**Table 2.** Electrical parameters of interface defects layer

| Parameters | CuI/Cs$_2$CuBiCl$_6$/CeO$_2$ |
|---|---|
| Defect Type | Neutral |
| Capture cross section for electrons/ hole (cm$^2$) | $1.00\times10^{15}$ |
| Energetic distribution | Single |
| Energy with respect to Reference (eV) | 0.6 Above the highest Ev |
| Characteristic energy /Ev | 0.1 |
| Total density (integrated over all energies) (cm$^{-2}$) | $1.00\times10^{15}$ |

**Table 3.** Previous and present work on different double perovskite layers

| Solar Cell Structures | Voc (V) | Jsc (mA/cm$^2$) | FF (%) | PCE (%) | References |
|---|---|---|---|---|---|
| FTO/TiO$_2$ /Cs$_2$CuBiCl$_6$ /Cu$_2$O | 0.91 | 21.66 | 85.99 | 17.03% | [12] |
| ITO/SnO$_2$/Cs$_2$AgBiBr$_6$/MoO$_3$ | 1.419 | 9.4741 | 72.61 | 11.41% | |
| FTO/ETL/Cs2AgBiBr6/HTLs/Cu | 7.2 | 8.02 | 6.45 | 3.75 % | [24] |
| FTO/SnO$_2$/Cs$_2$AgBiBr$_6$/Cu$_2$O/Au | 1.09 | 1.73 | 0.76 | 1.44% | [25] |
| FTO/ZnOS/Cs$_2$AgBi$_{0.75}$Sb$_{0.25}$Br$_6$ /Cu$_2$O | 1.39 | 16.04 | 78.34 | 18.8% | |
| FTO/ TiO$_2$ / Cs$_2$TiBr$_6$ /Cu$_2$O | 1.10 | 25.82 | 51.74 | 14.68% | [26] |
| FTO/ZnO/ Cs$_2$ AgBiBr$_6$/ NiO/Au | 1.29 | 20.69 | 81.72 | 21.88% | |
| FTO/CeO$_2$/ Cs$_2$CuBiCl$_6$/CuI/Au | 0.897 | 33.639 | 67.65 | 20.43% | **Before optimization** |
| FTO/CeO$_2$/ Cs$_2$CuBiCl$_6$/CuI/Au | 1.7359 | 32.829760 | 43.01 | 24.51% | **Final results** |

### 3. Results and Discussion

This research aims to explore the photovoltaic performance of Cs$_2$CuBiCl$_6$ (CCBIC) and also changes some parameters of the absorber layer (AL), which affect its efficiency. This work studies several factors that influence solar cell performance, including thickness, defect density (Nt), Electron affinity, band gap (Eg), and the system's temperature. By changing these factors, we also studied their impact on FF, Voc, PCE, and J$_{sc}$. During all parameter changes, the optimized parameter remains unchanged throughout the optimization process.

#### 3.1. Influence of the double perovskite layer (DPL) thicknesses

The absorber layer's thickness plays a crucial role in determining the performance of PSCs. When light falls on solar cells, electron-hole pairs are generated, leading to electricity production. An optimized PL thickness is crucial for maximizing the device's generation of charge carriers (electrons and holes). In this study, the PL thickness was varied from 0.450 to 2.000 µm while maintaining the other parameters of different layers, as presented in Tables 1 and 2. It was observed that increasing the PL thickness negatively impacted the key solar cell parameters, including the V$_{oc}$, J$_{sc}$, FF, and PCE, as illustrated in Figure 2 (a, b). This decline in performance can be attributed to increased charge recombination and light transmittance losses within the device at greater thickness. The absorber layer's thickness significantly affects the solar cell performance since both the saturation current density (Jo) and the device's short circuit current ($J_{sc}$) are close to open-circuit voltage (VOC). According to the equation 4:

$$Voc = \frac{KT}{q} \ln\left(\frac{jsc}{jo} + 1\right) \qquad 4$$

Where K is the Boltzmann constant, T is the temperature, and Q is the electronic charge, it is evident that any changes in Jo or Jsc due to absorber layer thickness directly influence Voc. Thus, optimizing the PL thickness is essential to achieving balanced charge generation and transport, minimizing recombination losses, and maximizing the overall device performance.

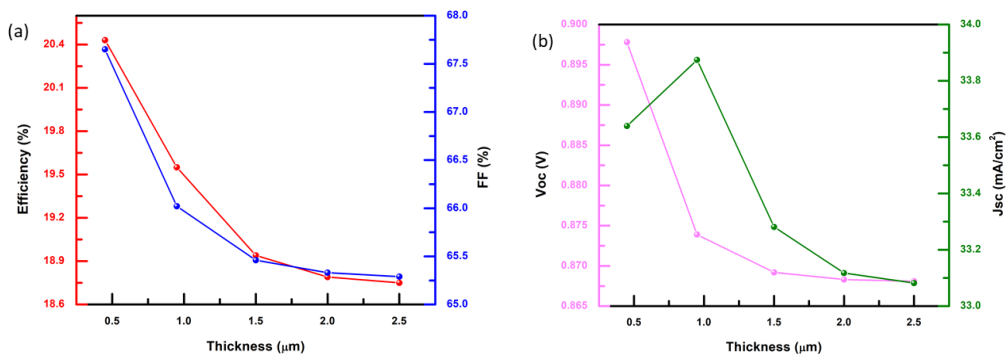

**Figure 2.** Impact of thickness on (a) FF and Efficiency (b) Voc and Jsc.

### 3.2. Influence of the DPL defect density ($N_t$)

In this study, the defect density (Nt) within the absorber layer (AL) was systematically varied over a range from 1E15 cm$^{-3}$ to 1E19 cm$^{-3}$ to evaluate its impact on the device performance of the solar cell. Optimizing the defect density is critical for enhancing device performance, as Nt directly influences the quality of the absorber film. An increase in defect density deteriorates the film quality, leading to a higher rate of charge carrier recombination, adversely affecting the solar cell's overall efficiency. A high defect density in the AL has been identified as a primary cause of reduced PCE due to the abundance of recombination centers [27, 28]. Figure 3(a, b) illustrates the trend in key performance parameters, including PCE, $J_{sc}$, and FF as a function of Nt. The results show that increasing Nt leads to a significant reduction in these parameters. However, variations in Nt did not result in noticeable fluctuations in the $V_{oc}$. One possible reason for this reduction in performance is the increased likelihood of charge carriers (electrons and holes) being trapped by defects in the material at higher Nt values. Consequently, at high defect densities, all key output metrics of the solar cell- PCE, $J_{sc}$, FF and $V_{oc}$ experience significant degradation. From the analysis, it was established that optimal device performance was achieved at a defect density of 1E15 cm$^{-3}$, which yielding notable values: PCE of 20.43%, $V_{oc}$ of 0.8978V, $J_{sc}$ of 33.639722mA/cm$^2$, and FF of 67.65% [34, 35]. The result underscores the critical role of minimizing defect density to achieve high-performance solar cells. Elevated defect densities introduce numerous recombination centers and traps, which impede charge carrier transport and significantly reduce device functionality. It is evident that superior device performance can only be attained by maintaining a low defect density in the absorber layer.

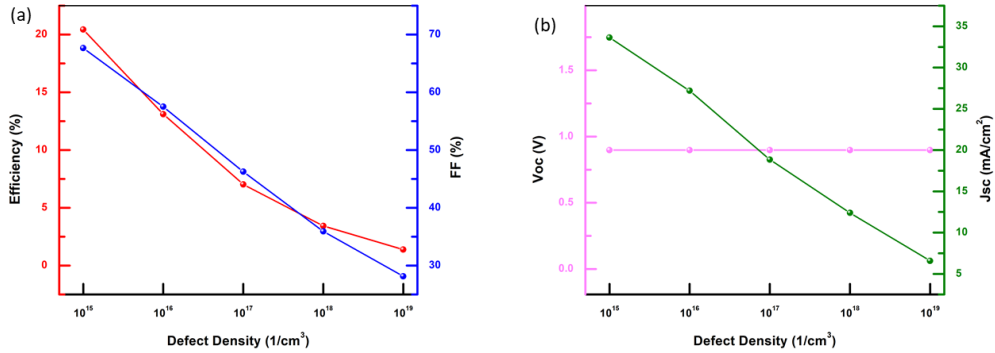

**Figure 3.** Impact of defect density on (a) FF and Efficiency (b) Voc and Jsc.

### 3.3. Influence of the DPL electron affinity

The electron affinity of the perovskite material (PM) is a critical factor that significantly impacts the performance of PSCs. In this study, the electron affinity of the absorber layer was optimized using SCAPS-1D simulations to evaluate its effect on the device's performance. The electron affinity of the PM determines the energy-level alignments between the absorber and the ETL and between the absorber and the HTL. Proper alignment is essential for ensuring efficient charge carrier transport and minimizing energy losses, directly affecting the overall device performance. To analyze the effect of electron affinity, the value of the absorber layer's electron affinity was systematically varied from 4.1 eV to 4.7 eV while keeping the $N_t$ of the perovskite layer constant at 1E15 cm$^{-3}$. Figure 4(a, b) presents the trends in all solar cell performance parameters as a function of the absorber's electron affinity. The results indicate that increasing the electron affinity initially enhances the device's performance. Specifically, the PCE improved from 20.43% at 4.1 eV to a maximum of 24.51% at 4.3 eV, marking a substantial enhancement in efficiency, as shown in Figure 4 (a). This improvement can be attributed to optimized energy-level alignment, which facilitates efficient charge carrier transport and minimizes recombination losses at the interfaces between absorber and transport layers. While the initial increase in electron affinity enhances charge transport efficiency, excessively high electron affinity values likely result in excessive electron transport from the ETL to the absorber, destabilizing the charge separation process. This destabilization negatively impacts overall device performance and leads to a notable reduction in PCE.

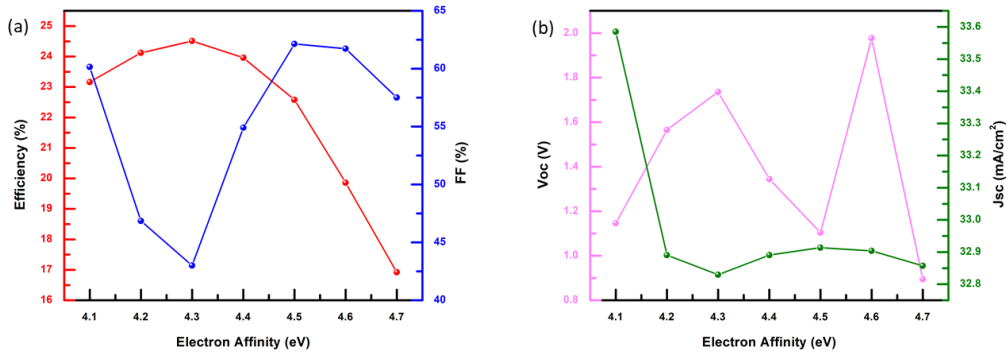

**Figure 4.** Impact of electron affinity on (a) FF and Efficiency (b) Voc and Jsc.

The $V_{oc}$, $J_{sc}$, and FF show a random fluctuation as the electron affinity increases, as shown in Figure 4 (a, b). While the initial increase in electron affinity enhances charge transport efficiency, excessively high electron affinity values likely result in excessive electron transport from the ETL to the absorber, destabilizing the charge separation process. This destabilization negatively impacts overall device performance and leads to a notable reduction in PCE. The results demonstrated that optimizing the electron affinity of the perovskite absorber layer is critical for achieving high device performance. An electron affinity value of approximately 4.3 eV is identified as an optimal condition, yielding the highest PCE of 24.51%. Beyond these values, the misalignment of energy levels between absorber and transport layers compromises charge carrier dynamics, reducing device efficiency. These findings highlight the importance of fine-tuning the electronic properties of the absorber layer to achieve the best possible performance in PSCs.

### 3.4. Influence of band gap (Eg) of the absorber layer

Numerous factors influence the efficiency and productivity of the photovoltaic cell performance. [29, 30]. Among these, the band gap of the absorber layer is a critical parameter. The band gap represents the energy difference between the valance band maximum and the conduction band minimum, directly impacting the light absorption, charge carrier generation, and overall performance of photovoltaic devices. This study systematically investigated the impact of varying the band gap of the absorber layer on the performance of PSCs. To evaluate this effect, the band gap of the absorber layer was varied from 1.12eV to 1.45 eV. Figure 5 (a,b) displays the I-V curve of the PSC with the configuration of $CeO_2$/$Cs_2CuBiCl_6$ /CuI as a function of the absorber layer band gap. The analysis reveals a clear trend; as the band gap of the perovskite layer increases, the PCE also increases, reaching a peak before declining at higher band gap values. The maximum PCE of 20.086% was achieved at an optimal band gap of 1.4 eV, corresponding to $V_{oc}$ of 1.26%, $J_{sc}$ = 54.74 mA/cm2, and FF = 54.74%. This enhancement in the PCE up to a band gap of 1.41 eV can be attributed to improved light absorption in the visible spectrum and better alignment of energy levels, which enhance charge carrier separation and transport. As the band gap increases beyond 1.41 eV, a noticeable decline in PCE is observed. At a band gap of 1.45 eV, both Jsc and FF significantly decrease, while Voc exhibits moderate increases. This reduction in Jsc is likely due to decreased light absorption, as a higher band gap limits the range of the solar spectrum that the absorber layer can effectively absorb. Similarly, the decline in FF can be attributed to increased recombination, adversely affecting charge carrier collection efficiency. The moderate increase in $V_{oc}$ with the band gap can be explained by the reduced radiative recombination rate and better separation of quasi-Fermi levels. However, the trade-off between $V_{oc}$ and improvement and Jsc reduction beyond the optimal gap results in an overall decline in device performance. This highlights the decline balance required in band gap optimization to achieve maximum efficiency. The results emphasize the critical role of the absorber layer's band gap in determining PSC performance

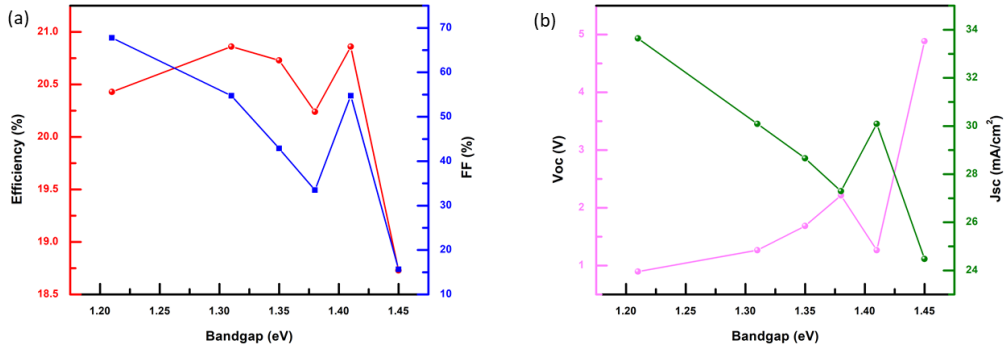

**Figure 5.** Impact of band gap on (a) FF and Efficiency (b) Voc and Jsc.

### 3.5. Influence of Operating Temperature

The operating temperature of PSCs has a significant impact on their efficiency and stability. Understanding these temperature-related effects is crucial for optimizing both the performance and durability of PSCs, particularly in outdoor applications where temperature fluctuations are frequent. This study's influence on temperature range varied from 300 k to 320 k [31]. Figure 6(a, b) illustrates the variation in key performance parameters, including Voc, Jsc, FF, and PCE as a function of temperature. The simulation results reveal that as the operating temperature increases, Jsc, PCE, and FF exhibit an increasing trend [32]. Specifically, the PCE enhancement with rising temperature can be attributed to factors such as alternations in material properties, including band gap, carrier transport concentration, and electron and hole mobilities. These temperature-induced changes align with analytical predictions and highlight the dynamic nature of PSC performance under varying thermal conditions. Interestingly, the decline in $V_{oc}$ with increasing temperature contrasts with the improvements observed in other parameters. This reduced Voc can be attributed to an increased density of interfacial defects, leading to higher series resistance and a decrease in carrier diffusion length. Elevated temperature tends to enhance thermal excitation, increasing interface recombination losses and adversely affecting the voltage output. The simulation also demonstrated that PSCs based on the $CS_2CuBiCl_6$ (CCBC) absorber layer exhibit promising performance at elevated temperatures. At 320 K, the PSC achieves a high PCE of 20.51%, $V_{oc}$ of 0.892 V, $J_{sc}$ of

33.71 mA/cm$^2$, and FF of 68.18%. These results underscore the potential of CCBC-based PSCs for application in environments with moderate temperature variations, as their performance parameters remain robust under such conditions. The findings highlight the critical importance of thermal stability and temperature optimization in PSC design.

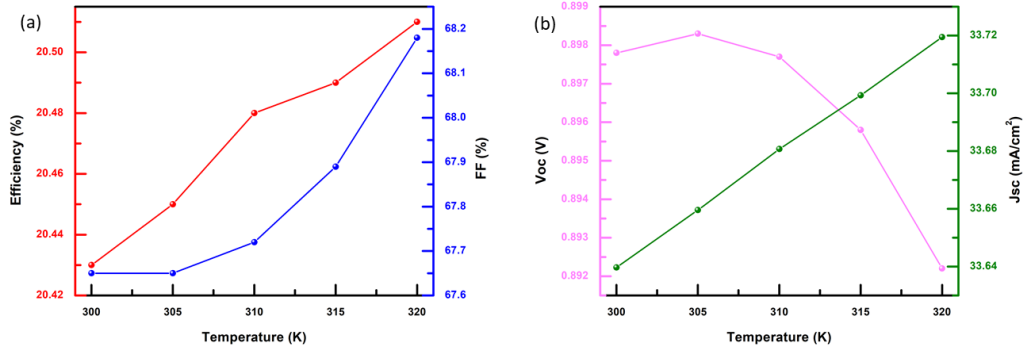

**Figure 6.** Impact of temperature on (a) FF and Efficiency (b) Voc and Jsc

### 4. Conclusion

This research presents a comprehensive investigation of Cs$_2$CuBiCl$_6$-based PSCs with an FTO/CeO$_2$/Cs$_2$CuBiCl$_6$/CuI/Au structure, focusing on the optimization of critical absorber layer parameters using SCAPS-1D simulations. Key parameters were systematically analyzed to enhance device performance, including thickness, defect density (Nt), electron affinity, band gap, and operating temperature. The findings demonstrated that optimizing the absorber layer's thickness enhances electron-hole pair generation and reduces recombination losses. Defect density emerged as a crucial determinant of film quality, with lower Nt 1E15 cm$^{-3}$ achieving a PCE of 20.43%. The electron affinity of the perovskite absorber was identified as a significant factor influencing energy-level alignment and charge transport. Band gap tuning revealed that an energy gap of 1.41 eV provided the highest efficiency of 20.86%. Furthermore, operating temperature was found to top impact device performance, with a PCE of 20.51% achieved at 320 K, $V_{oc}$ to 0.892 V, $J_{sc}$ to 33.71 mA/cm$^2$, and FF to 68.18%. The optimized cell performance, characterized by a maximum efficiency of 24.51%, highlights the potential of Cs$_2$CuBiCl$_6$ as a viable lead-free alternative in PSC applications. This work provides critical insight into the role of absorber layer parameters in device performance. It establishes a foundation for further experimental and theoretical advancements in lead-free photovoltaic materials.

**Data availability statement**

Data is available upon request.

**Declaration of competing interest**

The authors declare that they have no known competing financial interests or personal relationships that could have appeared to influence the work reported in this paper.